\documentstyle[twocolumn,letter]{jpsj2}

\title
{
Theory of Nonlinear Meissner Effect 
in High-$T_{\rm c}$ Superconductors}

\author
{
Takanobu {\sc Jujo}\thanks{E-mail: jujo@ms.aist-nara.ac.jp}
}

\inst
{
Graduate School of Materials Science, Nara Institute of Science 
and Technology, Ikoma, Nara 630-0101
}

\recdate
{\today
}

\abst
{
We investigate the nonlinear Meissner effect microscopically. 
Previous studies 
did not consider a certain type of interaction effect
on the nonlinear phenomena. 
The scattering amplitude barely appears without 
being renormalized into the Fermi-liquid parameter. 
With this effect we can 
solve the outstanding issues 
(the quantitative problem, 
the temperature and angle dependences). 
The quantitative calculation is performed with use of the fluctuation-exchange 
approximation on the Hubbard model. 
It is also shown that the perturbation expansion on the 
supercurrent by the vector potential converges owing 
to the nonlocal effect. 
}

\kword
{
nonlinear Meissner effect, unconventional superconductor, 
electron correlation, spin fluctuation, 
vertex correction, nonlocal effect
}

\begin{document}
\maketitle

Many unconventional superconductors have recently been discovered. 
The evidence that these superconductors are 
non-s-wave is obtained by 
thermodynamic measurements in most cases. 
These types of measurement provide only 
information averaged over the Fermi surface 
and then other measurements are needed 
to determine the position of nodes. 
Detailed information about the node 
is useful for judging the accuracy of 
various theories. 
One of these, the nonlinear Meissner effect (NLME), 
which provides a measurement of 
the magnetic field ($H$) dependence of the magnetic field 
penetration depth ($\lambda+\delta \lambda(H)$), 
was proposed by Yip and coworkers (YS).~\cite{yip,xu} 
Their proposal is based on the Doppler-shifted energy spectrum 
and its predictions are summarized as follows. 
(i) The supercurrent has a nonanalytical form 
(written as $A|A|$, $A$ is the vector potential) and 
$\delta \lambda(H)$ is proportional to the magnetic field 
$|H|$ as a result. 
(ii) $\delta \lambda(H)$ varies with 
the direction of the applied magnetic field and therefore 
$\delta \lambda(H)$ provides information on the position 
of nodes. 

To date, the experiments have not provided decisive 
results because the effect is small and 
tends to be masked by many extrinsic effects. 
The first investigation of NLME was carried out by 
Maeda {\it et al.}~\cite{maeda}, however this 
experiment was performed with the magnetic field 
perpendicular to the CuO plane. 
The precision was also poor (order of 10\AA) 
and, at present, the observed quantity is considered to reflect 
extrinsic effects. 
Experiments with high precision (order of 0.1\AA) 
were carried out by Bidinosti {\it et al.}~\cite{bidinosti} and 
Carrington {\it et al.}~\cite{carrington} 
The most reliable results in ref.4 
are summarized as 
$\delta \lambda (H) \propto H^2$, the temperature dependence 
is weak and the angle dependence is not observed. 
These contradict YS's theory. 
YS also predict that 
$\delta \lambda(H) \propto H^2$ below 
the crossover field, however, 
the temperature dependence is strong. 
Two groups have attempted to detect the transverse 
magnetization, which is the other prediction 
made by YS.~\cite{bhattacharya,buan}  
The recent experiment with higher precision 
(two orders of magnitude)~\cite{bhattacharya} 
showed that the amplitude of this quantity is at most 
one third of the predicted one and it is almost 
at the measurable limit. 
Extensions of YS's theory were made 
by several researchers.~\cite{stojkovic,li,dahm,halterman}  
The common results obtained by them are that the 
theoretical predictions are inconsistent with 
the experimental results. 
Therefore, some papers 
suggest that the experiments observed extrinsic effects. 
(In ref.5 
the possible extrinsic effects are listed 
(the vortex contribution, the weak links, 
the interlayer contribution).)
The important point is that 
the values predicted by YS and other researchers are 
at least of the same order of magnitude, compared with the 
values obtained in the experiments. 
Therefore, it is not possible that the intrinsic theoretical 
value is masked by the extrinsic effects which are suppressed 
up to 0.2\AA. 

Then, the question arises as to whether 
the existing theories are correct for judging 
experimental results. 
Here, we discuss the NLME effect 
on the basis of the perturbation theory and 
show that the previous theories have some defects. 
The perturbation expansion by the vector potential 
on the supercurrent and the magnetic field penetration depth converges 
owing to the nonlocal effect. 
The intermediate-states interaction~\cite{comment1} (electron-electron) 
which is not included in the conventional 
quasiclassical approximation~\cite{comment2} 
exists and makes a dominant contribution. 
This effect solves the inconsistency 
between the theory and the experiments on the value 
of $\delta \lambda$ and its angle and temperature dependences. 
We adopt the fluctuation-exchange (FLEX) approximation 
for the quantitative calculation. 
The many-body effect on the response function is included 
on the basis of the conserving approximation. 

Our theory is based on the evaluation of the response function 
in the supercurrent which is exactly expanded by $A$ up to the third order. 
The expression for the supercurrent is written as 
\begin{eqnarray}
&&\hspace{-1.0cm} J_{\mu}(q)= 
-K^{(1)}_{\mu\nu}(q)A_{\nu}(q) 
-\int_{q'}K^{(2)}_{\mu\nu\alpha}(q,q')A_{\alpha}(q')A_{\nu}(q-q') \nonumber\\
-&&\hspace{-0.7cm} 
\int_{q',q''}K^{(3)}_{\mu\nu\alpha\beta}(q,q',q'')
A_{\beta}(q'')A_{\alpha}(q'-q'')A_{\nu}(q-q'), 
\end{eqnarray}
where $K^{(1,2,3)}$ are the response functions 
in the perturbation expansion, 
$\mu,\nu,...$ are the spatial dimensions and 
the summation of the repeated indices is taken. 
$K^{(1)}$ appears in the usual linear response theory. 
The $K^{(3)}$ term is dominant in the 
magnetic field dependence of $\lambda$ because 
$K^{(2)}$ vanishes. 

First, we show the convergence of the perturbation expansion. 
By analyzing various terms it is shown that the most 
divergent term in the local limit 
is the type (a) term in Fig. 1 and is written as, 
$K^{(3a)}(q,q',q'') 
=\frac{T}{V}\sum_{\mib k,n}v^{4}_{\mib k}{\rm Tr}
[\hat{G}_{k+q/2}\hat{G}_{k+q'-q/2}\hat{G}_{k+q''-q/2}\hat{G}_{k-q/2}]$, 
where $T$, $v_k$ and 
$\hat{G}_k$
are the temperature, the velocity and 
Green's function in the superconducting state, 
respectively, and 
$k=(\mib k,{\rm i}\epsilon_n)$ ($\epsilon_n=\pi T (2n+1)$, 
$n$ is integer) and $q=(\mib q,0)$. 
The uniform component is written as 
$K^{(3a)}(0,0,0) \propto 
-\frac{v}{\Delta_0}\frac{1}{T}$ 
($\Delta_0$ is the maximum of the superconducting gap). 
This is the result of YS's theory. 
$K^{(3)}(q,q',q'')$ diverges for $T\rightarrow 0$ at $q=q'=q''=0$, 
however, this term is integrated by $q'$ and $q''$ 
in the expression of the supercurrent and by $q$, $q'$ and $q''$ 
in the case of $\delta \lambda$. 
Then, the behavior of $K^{(3)}(q,q',q'')$ 
in $q$-space comes into question. 
The $q$-dependence of $K^{(3a)}(q,0,0)$ for small $q$ and at $T=0$ 
is written as 
$K^{(3a)}(q,0,0) \propto 
-\frac{v}{\Delta_0}\frac{1}{v_{\perp}|q|}$ 
($v_{\perp}$ is the mean value of the the interlayer velocity 
with magnetic field parallel to the $ab$-plane.) 
It is difficult to determine analytically the dependence 
of $K^{(3)}(q,q',q'')$ on $q$, $q'$ and $q''$, 
however the nonzero values of $q'$ and $q''$ do not make this term 
more divergent than $1/q$ and if we consider the symmetry relation, 
$K^{(3)}(q,q,q)=K^{(3)}(q,0,0)$, the form for small $q$, $q'$, $q''$ 
is considered to be, 
$K^{(3)}(q,q',q'') \propto 1/\sqrt{
(\frac{q}{2})^2+(q'-\frac{q}{2})^2+(q''-\frac{q}{2})^2+(\frac{q''}{2})^2}$. 
The $q^{-1}$ divergence of $K^{(3)}(q,q',q'')$ 
guarantees the convergence of the perturbation expansion on 
the physical quantities. 
Therefore, the nonanalytical behavior does not appear 
contrary to the prediction of YS.~\cite{comment3} 

\begin{figure}
\includegraphics[width=4cm]{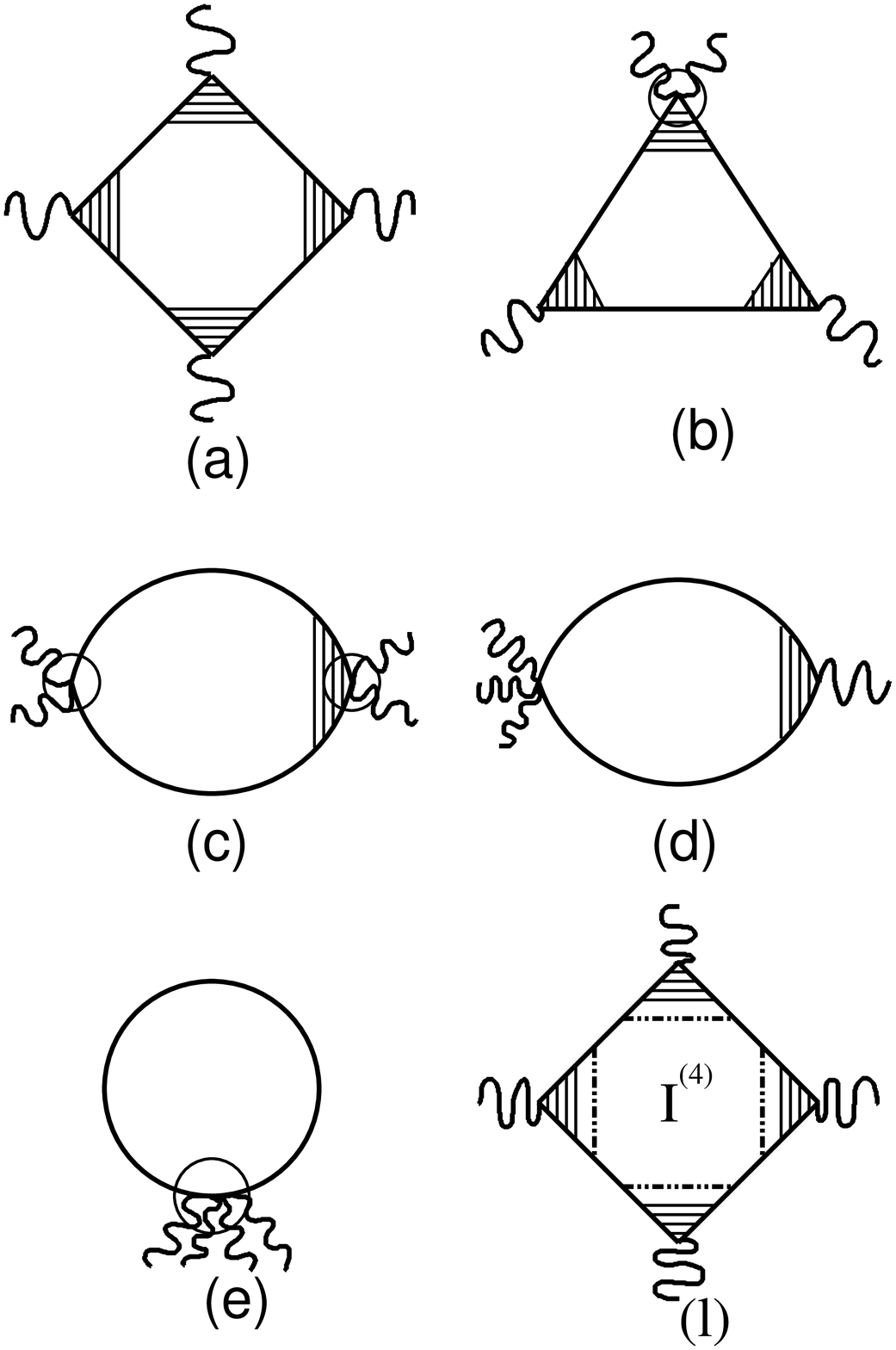}
\includegraphics[width=4.3cm]{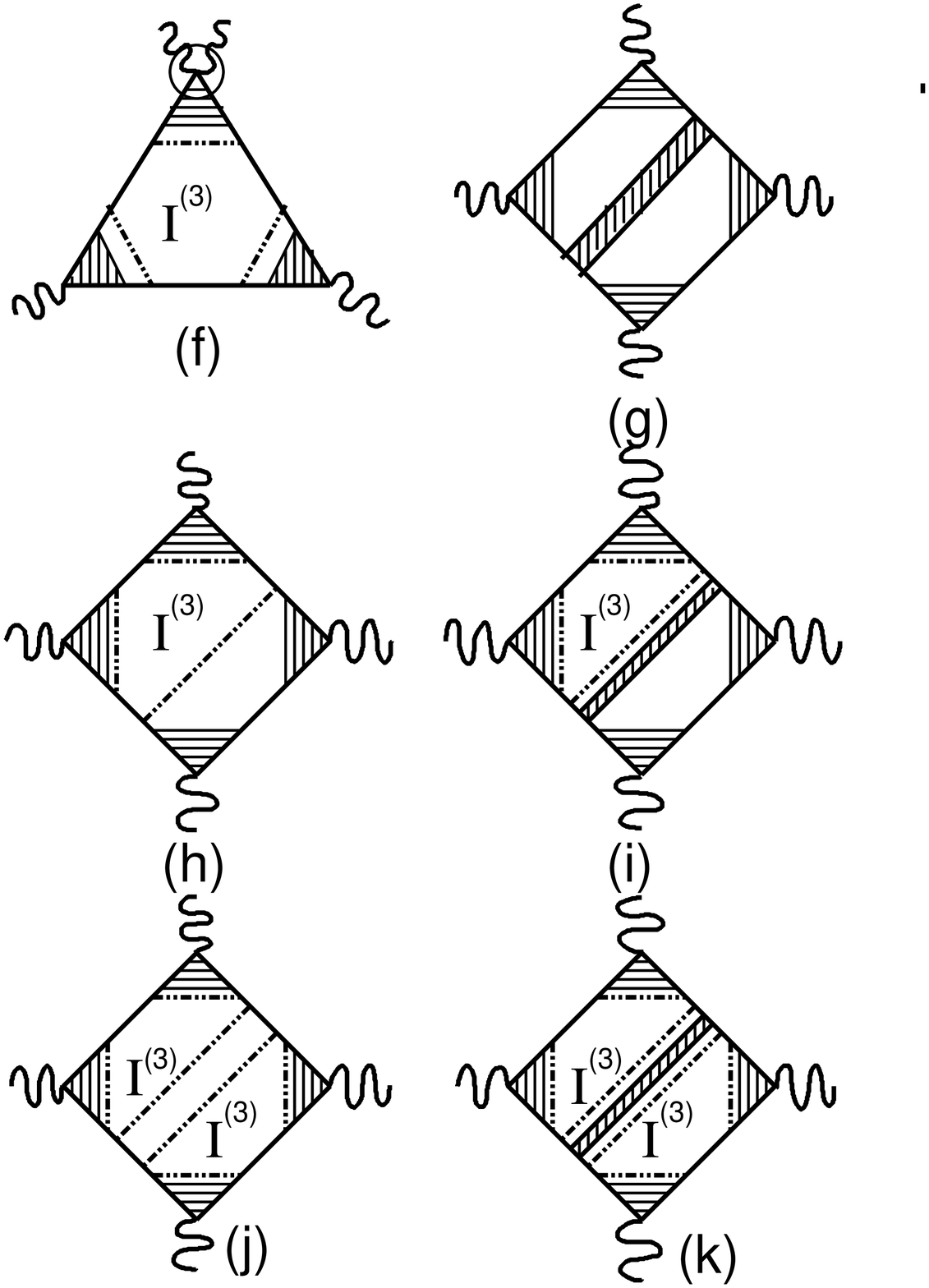}
\caption{Diagrammatic representation of $K^{(3)}$.
The solid lines express the propagator of the electron 
in the Nambu representation. 
The wavy lines express the electro magnetic fields. 
The vertex with $n$-wavy lines denotes the $n$-times derivative 
of the dispersion of electrons and the circle at the vertex means 
a diagonal matrix $\hat{\tau}_3$ with ${\rm Tr}\hat{\tau}_3=0$. 
The vertex with a shaded triangle satisfies the integral equation 
with a four-point irreducible vertex $I^{(2)}$. 
The shaded rectangle denotes the reducible 
four-point vertex $\Gamma^{(2)}$. 
$I^{(3)}$ and $I^{(4)}$ represent the irreducible 
six- and eight- point vertices, respectively. 
}
\label{fig:1}
\end{figure}

Next, we estimate the various terms in $K^{(3)}$. 
(The diagrammatic representation of $K^{(3)}$ is 
given in Fig.~\ref{fig:1}.)
There are many cumbersome terms unlike the linear 
response case.~\cite{comment4} 
The approximation used is as follows. 
The three-point vertex correction connected with the odd order 
of the external field is omitted. 
This is  because the velocity is an odd function 
in wave-number space and the integral is small. 
The same approximation holds in some types of 
the six- and eight-point irreducible vertices. 
If we consider a system with a strong momentum dependence 
such as the underdoped region and its doping dependence, 
this type of vertex is necessary.~\cite{jujo} 
We consider, however, mainly the temperature and the 
angle dependence, and then 
the above terms have a slight influence. 

The formalism for deriving $\delta \lambda(H)$ consists 
of the Maxwell equation with the specular boundary condition 
(here, we consider the situation where the external magnetic 
field is applied parallel to the $y$-axis and the superconductor 
occupies $z>0$) $\frac{{\rm d}^2A_{x}(z)}{{\rm d}z^2}=
2H \delta(z)-\frac{4\pi}{c}J_{x}(z)$, 
($H$ is the applied external field) 
and the nonlinear Pippard equation (eq. (1))
From these two equations the nonlinear equation for $A$ 
is obtained and is solved by the perturbation method 
(not self-consistently) because we use the perturbation 
method to obtain the response kernel $K^{(3)}$. 
Then, the nonlinear correction for $\lambda$ is obtained as 
(the magnetic field is parallel to the intralayer crystal axis) 

\begin{eqnarray}
\delta \lambda_{ab}&=&-8H^2 \frac{4\pi}{c}
\int\frac{{\rm d}q}{2\pi}\int\frac{{\rm d}q'}{2\pi}\int\frac{{\rm d}q''}{2\pi} 
K^{(3)}_{xxxx}(q,q',q'') \nonumber\\
&& \times 
D_{xx}(q'')D_{xx}(q'-q'')D_{xx}(q-q')D_{xx}(q), 
\end{eqnarray}
where the definition of the magnetic field penetration depth is 
$\lambda:=\frac{1}{H}\int_{0}^{\infty}H_{y}(z){\rm d}z
=-\frac{1}{H}\int\frac{{\rm d}q}{2\pi} A_x(q)$, and 
$D_{xx}(q):=1/(q^2+\frac{4\pi}{c}K^{(1)}_{xx}(q))$. 

First, we consider the terms without the intermediate-states 
interaction. 
We classify these terms into two groups. 
One group consists of divergent terms without nonlocality and 
the other of terms similar to $K^{(1)}$ 
(the nonlocality is negligible). 
$K^{(3a)}$ and $K^{(3b,c,d,e)}$ 
are categorized into the former and the latter groups, respectively. 
We estimate these terms only with the experimentally observed values. 
(On the other hand, a specific microscopic model and 
approximation are needed in the case of the intermediate-states 
interaction.) 
It is difficult to analytically calculate 
the paramagnetic term $K^{(3a)}$ 
at a finite temperature, however, by noting 
that $E_{k+q}-E_k \simeq v_{\perp}q$ near nodes and the characteristic 
value of $q$ is $\lambda^{-1}$ it can be shown that there is 
a crossover temperature 
$T_0 \simeq \frac{\hbar v_{\perp}}{\lambda} 
\simeq \frac{\xi_c}{\lambda}\Delta_0$ 
($\xi_c$ is the interlayer coherence length). 
By substituting $K^{(3a)}(q,q',q'')$ into eq. (2), 
$\delta \lambda^{(3a)}
=\left(\frac{H}{H_c}\right)^2 \lambda
\kappa^{(3a)}(\frac{T_0}{\Delta_0})$. 
Here, $\kappa^{(3a)}(\frac{T_0}{\Delta_0})$ is a dimensionless 
quantity and its integration is calculated numerically. 
($H_c:=\phi_0/2\sqrt{2}\pi\xi_{ab}\lambda$ is the 
thermodynamic critical field, $\phi_0=2\pi \hbar {\rm c}/2{\rm e}$, 
$\xi_{ab}$ is the in-plane coherence length). 
$\kappa^{(3a)}(\frac{T_0}{\Delta_0})$ is strongly 
dependent on temperature 
and shows a maximum around $T\simeq T_0$. 
If we put 
$\lambda=1600$\AA, 
$T_0/\Delta_0=0.01$ and $H_c=8000$G, 
$\delta \lambda^{(a)} \lesssim 0.2{\rm \AA}$ 
for $H=200$G; this is small compared with the experimental result 
$\delta \lambda|_{\rm exp.}\simeq 5 {\rm \AA}$. 
$\delta \lambda^{(b,c,d,e)}$ is smaller than one tenth of 
$\delta \lambda$ due to the intermediate-states interaction 
discussed below and is negligible. 

Next, we consider the intermediate-states interaction. 
(The local approximation holds as in the case of (b,c,d,e) 
and $K^{(1)}$.)
The terms are calculated after they are transformed 
into equations which explicitly equal 0 when $\Delta=0$. 
The term with $\Gamma^{(2)}$ ((g)-type) is evaluated as follows. 
The reducible four-point vertex satisfies 
\begin{equation}
\breve{\Gamma}^{(2)}(k,k')
=\breve{I}^{(2)}(k,k')
+\frac{T}{V}\sum_{k''}\breve{I}^{(2)}(k,k'')\breve{g}(k'')
\breve{\Gamma}^{(2)}(k'',k')
\end{equation}
with the irreducible four-point vertex $\breve{I}^{(2)}(k,k')$. 
($\breve{}$ denotes the $4\times 4$ matrix in the particle-hole space.)
We use eq. (24) in ref.17 
as $\breve{g}(k)$. 
There are various terms in $\breve{I}^{(2)}(k,k')$, 
the particle-hole vertex, the particle-particle vertex, 
the number-nonconserving vertices. 
These are the functional derivatives 
of the self-energy by Green's function 
(the conserving approximation~\cite{baym}). 
One of these vertices (the number-conserving particle-hole vertex) 
is written as follows with the FLEX approximation 
(the FLEX approximation in the superconducting state, 
for example, see refs.19 and 20) 
\begin{equation}
I_c(k,k')=V^n_{k-k'}
-\frac{T}{V}\sum_q G_{k-q}W_q(G_{k'-q}+G_{k'+q}). 
\end{equation}
Here, 
\begin{equation}
V^{n}_q=U
\left[U\chi^s_q
+\frac{3}{2}\frac{(U\chi^s_q)^2}{1-U\chi^s_q}
-\frac{1}{2}\frac{(U\chi^c_q)^2}{1+U\chi^c_q}
\right], 
\end{equation} 
with 
$\chi_q^{s,c}=-\frac{T}{V}\sum_k(G_{k+q}G_k \pm F_{k+q}F_k)$, 
($(s,c)$ correspond to $(+,-)$,
and 
\begin{equation}
W_q=U^2
\left[
\frac{3}{2}\frac{1}{(1-U\chi^s_q)^2}
+\frac{1}{2}\frac{1}{(1+U\chi^c_q)^2}-1
\right].
\end{equation} 
The other vertices are derived in the same way. 
We adopt the Hubbard model with the on-site Coulomb interaction $U$ 
and take the same dispersion of electrons as in ref.16. 
The terms with $I^{(3)}$ and $I^{(4)}$
are calculated in the same way 
(e.g., $I^{(3)}(k_1,k_2,k_3)=
\delta^2 \hat{\Sigma}_{k_1}/\delta \hat{G}_{k_3}\delta \hat{G}_{k_2}$). 
We calculate $K^{(3)}$ without integrating out the 
incoherent part to derive 
a low-energy expression (e.g., ref.16.)
Therefore, the effect of the renormalization factor 
is implicitly included. 

To compare the calculation with the experimental results quantitatively, 
we consider the following quantity 
\begin{equation} 
\frac{1}{H^2}\frac{\delta \lambda(H)}{\lambda}
=-\frac{\pi}{16}\frac{1/\phi_0^2}
{({\rm e}/{\hbar \rm c})^2K^{(1)}t/a}\frac{a^2 K^{(3)}}{K^{(1)}}.
\end{equation}
Here, $a$ is the lattice constant. 
We numerically calculate $K^{(3)}$ and $K^{(1)}$ by putting 
$t=1$ and $a=1$ and 
quantify the values of $t$ and $a$ in units of 
$[{\rm eV}]$ and $[{\rm \AA}]$, respectively, and then 
$\frac{1}{H^2}\frac{\delta \lambda(H)}{\lambda}
=-0.13\times 10^{-13} \times \frac{a^3}{t}
\frac{K^{(3)}}{(K^{(1)})^2} [{\rm G}^{-2}]$. 
If we put $t=0.25[{\rm eV}]$ and $a=8[{\rm \AA}]$, 
we get $\frac{1}{H^2}\frac{\delta \lambda(H)}{\lambda}
\simeq 0.35 \times 10^{-7} [{\rm G}^{-2}]$ for 
$U=6.0$ and the hole doping $\delta=0.20$. 
($K^{(3)}\simeq -12.8$, $K^{(1)}\simeq 0.1$ and 
the dominant contribution comes from $K^{(3g)}$. 
This value of $K^{(1)}$ yields $\lambda \simeq 2600 {\rm \AA}$. 
This is roughly 1.6 times longer than the experimental value.)
On the other hand, the experimental result is 
$\frac{1}{H^2}\frac{\delta \lambda(H)}{\lambda}
\simeq (0.7 \sim 1.0)\times10^{-7} [{\rm G}^{-2}]$~\cite{bidinosti}. 
Our calculation is quantitatively consistent with 
the experimental results in order of magnitude. 
As for the parameter dependence, the value of 
$\frac{1}{H^2}\frac{\delta \lambda(H)}{\lambda}$ 
is not strongly dependent on the parameters $U$ and $\delta$ 
in the FLEX calculation. 
For example, $\frac{1}{H^2}\frac{\delta \lambda(H)}{\lambda}
\simeq 0.34 \times 10^{-7} [{\rm G}^{-2}]$ for $U=7.0$ and 
$\delta=0.20$ and $0.33 \times 10^{-7} [{\rm G}^{-2}]$ 
for $U=6.0$ and $\delta=0.15$. 
This is because the effect of the renormalization factor on 
$K^{(3)}$ and $(K^{(1)})^2$ cancels each other (as for $U$-dependence) 
and the integral-equation structure for $\Gamma^{(2)}$ 
weakens the variation of the spin-fluctuation effect on $I^{(2)}$. 
If we put $W_q \rightarrow U^2$, this corresponds to 
the case of the weak spin fluctuation, for example, the 
more overdoped region, 
this results in a smaller value of 
$\frac{1}{H^2}\frac{\delta \lambda(H)}{\lambda}$. 
Therefore, an experimental study on the doping dependence 
is expected. 

To investigate the angle dependence we consider the case 
where the applied field is parallel to the 
node direction ($\delta \lambda_{45^{\circ}}(H)$). 
In this case, $K^{(3)}_{xxxx}$ in eq.(2) is 
replaced by $(K^{(3)}_{xxxx}+3K^{(3)}_{xxyy})/2$. 
Then the relationship between $K^{(3)}_{\mu\mu\mu\mu}$ and 
$K^{(3)}_{\mu\mu\alpha\alpha}$ with $\mu \ne \alpha$ 
plays an important role in the angle dependence. 
If we consider a conventional $s$-wave superconductor, 
the relation $K^{(3)}_{\mu\mu\mu\mu}=
3K^{(3)}_{\mu\mu\alpha\alpha}|_{\mu \ne \alpha}$ holds 
because 
$<v^4_{\mu}>_{\rm FS}=3<v^2_{\mu}v^2_{\alpha}>_{\rm FS}$ 
in the superconductor with no nodes and 
$v_{\mu\mu\mu}=0$ and 
$v_{\mu\mu}^2=v_{\mu\mu}v_{\alpha\alpha}$ hold 
in the electron gas. 
($<...>_{\rm FS}$ denotes the average over the Fermi surface 
and $v_{\mu\mu}=\partial v_{\mu}/\partial k_{\mu}$, etc.) 
Both of these relations do not hold in the unconventional 
superconductor in the lattice system and therefore 
the relationship between $K^{(3)}_{\mu\mu\mu\mu}$ and 
$K^{(3)}_{\mu\mu\alpha\alpha}|_{\mu \ne \alpha}$ 
is not trivial. 
In fact, $K^{(3a)}_{xxxx}$ and $K^{(3a)}_{xxyy}$ give 
the same contribution to $\delta \lambda$ 
because a dominant contribution to the integral 
over the Fermi surface comes from nodes 
($v_x=v_y$ at this point) except for $T>>T_0$. 
Then, $\delta \lambda_{45^{\circ}}(H)=2\delta \lambda_{ab}(H)$ 
in the conventional quasi-classical approximation.~\cite{comment5} 
We made sure above, however, that the intermediate-states 
interaction term contributes to $\delta \lambda$ sufficiently 
and can be dominant. In this case, 
$K^{(3)}_{xxyy}=K^{(3)}_{xxxx}/3$ with the approximation noted above 
and then 
$\delta \lambda_{45^{\circ}}(H)=\delta \lambda_{ab}(H)$. 
This explains the experimental results. 
(The reason for this is that the correlation 
between different vertices is broken by 
the intermediate-states interaction. 
For example, 
$\int_q[\int_k G_k v_{k\mu} G_k v_{k\nu} G_k G_{k-q}
W_q \int_{k'} G_{k'-q} G_{k'} v_{k'\alpha} G_{k'} v_{k'\beta} G_{k'}]$
is negligible for $\mu \ne \nu$ or $\alpha \ne \beta$. 
The same holds for the case of $I^{(3)}$ and $I^{(4)}$. 
This discussion also applies to 
the temperature dependence.) 

The temperature dependence of $\delta \lambda$ is as follows. 
$\delta \lambda \propto 1/T$ for $T>T_0$ 
in the conventional quasi-classical approximation. 
On the other hand, the temperature dependence of 
the intermediate-states interaction term 
is same as that of $K^{(1)}$ ($T$-linear) and 
the decreasing rate compared with the value at $T=0$ 
is almost same. 
Therefore, $\delta \lambda$ shows a slight increase. 
(If $\lambda$ increases $5{\rm \AA}$ per $1{\rm K}$ 
as the experimental results indicate, the increasing rate of 
$\delta \lambda$ is $\frac{3\times 5}{\lambda/\delta \lambda} 
\simeq 0.015$ $[{\rm \AA}/{\rm K}]$.
This value is no larger than the experimental precision.)

The other phenomenon related to the NLME 
is the transverse magnetization. 
The predicted behavior in refs.1 and 2 
is that the supercurrent is not perpendicular 
to $H$ except for the case in which 
$H$ is parallel to the nodes or the antinodes, 
and therefore the transverse magnetization 
has a period of $\pi/2$ as the direction of $H$ 
is rotated. 
Our perturbation theory shows that 
the supercurrent is written as 
follows in the arbitrary direction of $H$. 
\begin{eqnarray}
&&J_{\mu}(q)
\simeq \{
K^{(1)}_{\mu\mu}(q)A(q) 
+\int_{q'}\int_{q''}
[K^{(3)}_{\mu\mu\mu\mu}(q,q',q'')X_{\theta}^2 \nonumber\\
&&+3 K^{(3)}_{xxyy}(q,q',q'')Y_{\theta}^2
]A(q'')A(q'-q'')A(q-q')
\}X_{\theta}. \nonumber\\
\end{eqnarray}
Here, $(X_{\theta},Y_{\theta})=({\rm cos}\theta,{\rm sin}\theta), 
({\rm sin}\theta,{\rm cos}\theta)$ 
for $\mu=x,y$, respectively. 
$\theta$ is the angle between the applied field and 
the intralayer crystal axis.)
If the relation 
$K^{(3)}_{\mu\mu\mu\mu}=
3K^{(3)}_{\mu\mu\alpha\alpha}|_{\mu \ne \alpha}$ holds, 
the transverse magnetization does not appear. 
Then, we can have the same discussion as in the case of $\delta \lambda$. 

Finally, we comment on previous studies. 
The nonlocal effect considered in ref.9 
is different from our approach in several points. 
The behavior $\delta \lambda \propto H^2$ 
at low $H$ is seemingly the same as that in the perturbation approach. 
They predict, however, $\delta \lambda \propto H$ 
above the crossover field $H^*$ and argue that 
the NLME is unobservable 
owing to $H^*>H_{c1}$. 
They consider that the nonanalytical current exists above $H^*$. 
Although they do not consider the angle dependence of $\delta \lambda$ 
and the transverse magnetization, 
their theory contradicts the experimental results. 
The origin of their error is that they consider 
$K^{(1)}(q,A_{q=0})$. 
They derive $H^*$ by comparing the effect of $q$ and $A$, 
however, it does not make sense to 
compare the intrinsic spatial variation with 
the external field. 
The absence (or very small value) of 
the transverse magnetization below the first 
vortex penetration~\cite{bhattacharya2} implies 
the absence of $H^*$. 

The quasi-classical approach in ref.23 
gives observable values 
($\delta \lambda \simeq 1 \AA$ for $H \simeq 200[G]$) 
with the experimental parameters 
($H^* \simeq 2[{\rm T}]$ and $\lambda/\xi \simeq 100$). 
Then, this also contradicts the experimental results 
in the angle and temperature dependences qualitatively. 
The cause is as follows. 
The interaction with the external field in the Gor'kov equation 
$\frac{1}{2m}\left(
\nabla_r+\frac{\nabla_R}{2}-{\rm i}\frac{\rm e}{\rm c}A(R+r/2)
\right)^2G(r,R)$
is approximated as 
$\frac{1}{2m}\left(
\nabla_r+\frac{\nabla_R}{2}-{\rm i}\frac{\rm e}{\rm c}A(R)
\right)^2G(r,R)$ 
in the quasi-classical approach. 
(The propagator is transformed as 
$G(x,x')\rightarrow G(x-x',\frac{x+x'}{2})=G(r,R)$.) 
This means that the external field interacts with the center of mass 
of the electron propagator and therefore 
the nonlocal effect is underestimated.  
The comparison with our $(a)$-term is as follows. 
The Green function in the third order of the external field 
is written as 
\begin{eqnarray}
G^{(3)}(k,q)&\rightarrow& \sum_{q_1,q_2}G_{k+q/2}vA_{q_2}
G_{k+q/2-q_2}vA_{q_1-q_2} \nonumber\\
&& \times G_{k+q/2-q_1}vA_{q-q_1}G_{k-q/2}
\end{eqnarray}
for the case of the nonlocal effect included correctly, but 
\begin{eqnarray}
G^{(3)}(k,q)&\rightarrow& \sum_{q_1,q_2}G_{k+q/2}vA_{q_2}
G_{k+q/2-q_2/2}vA_{q_1-q_2} \nonumber\\
&& \times G_{k+q/2-q_1/2}vA_{q-q_1}G_{k}
\end{eqnarray}
in the quasi-classical approximation. 
This is interpreted as meaning that 
the magnetic field penetration depth effectively 
doubles in this approximation and then $\delta \lambda$ 
roughly increases eightfold. 
Therefore, the quasi-classical term ($(a)$-term in our paper) 
makes less contribution to $\delta \lambda$ if it is evaluated 
properly. 

Larkin and Ovchinnikov suggest that 
the quasi-classical approximation does not 
give correct results in some cases.~\cite{larkin}
Our theory presents a definite example 
of this proposition.

In this paper, we present the microscopic formulation of the 
nonlinear Meissner effect. 
We show that the previous studies on this effect are 
insufficient and some of them are incorrect. 
The nonanalytical response is intrinsically absent. 
The experimental results possibly observe the 
intrinsic NLME. 
This is not YS's one, but originates from 
the intermediate-states interaction. 
We consider that this effect is interesting because 
it does not appear in the zeroth order of interactions 
but it reflects interactions between quasiparticles themselves. 
The spin fluctuation is quantitatively dominant in our 
calculation. 
This is consistent with the properties of 
the high-$T_c$ cuprates. 
Experiments on other materials and 
the theoretical investigations of various scattering mechanisms 
are expected in the future. 

Numerical computation in this work was carried out at 
the Yukawa Institute Computer Facility.


\begin{thebibliography}{99}
\bibitem{comment1} The intermediate-states interaction 
is termed in analogy with the final-state interaction. 
The latter term is well known, for example, in the exciton formation 
in the linear absorption spectrum. 
This effect arises from an interaction 
between the initial and final states. 
However, interactions between various states come into 
question with consideration of the nonlinear effect. 
This is the reason why we use this term.
\bibitem{comment2} The conventional quasi-classical approximation 
means the approximation which fails to treat the 
intermediate-states interaction properly, though 
an extension to include this effect remains an issue. 
The previous studies based on the quasi-classical approximation 
fall into this category.
\bibitem{comment3} 
Kosztin and Leggett 
[
I. Kosztin and A. J. Leggett: 
Phys. Rev. Lett. {\bf 79} (1997) 135. 
] 
show that the leading nonlocal term 
in the linear response kernel $K^{(1)}(q)$ is $|q|$-linear 
in the superconductors with nodes. 
Although it is not written in their paper, 
the perturbation expansion of $K^{(1)}(q)$ by $q$ 
diverges for $T\rightarrow 0$. 
This is the reason why the non-analytic term ($|q|$)
appears in the $q$-dependence of $K^{(1)}(q)$. 
YS's nonanalytical term is different from this 
because the external field ($A$) itself has the nonlocality. 
Therefore, the replacement of $q$ by $A$ causes 
an incorrect result.
\bibitem{comment4} 
Only (a) is taken in the conventional quasi-classical approximation. 
This corresponds to 
the expansion of only the paramagnetic term 
(the expansion of the Fermi distribution function by $A$). 
This brings about the violation of the gauge invariance.
\bibitem{comment5}
Furthermore the quantum correction term ($K^{(3b,c,d,e)}$)
which has not been considered 
in previous studies makes the angle dependence of $\delta \lambda(H)$ 
nontrivial and dependent on the properties of the particular system.
\end{thebibliography}
\end{document}